\documentclass[utf8]{frontiersSCNS} 

\usepackage[onehalfspacing]{setspace}

\newcommand{\apj}{Astrophys. J. }
\newcommand{\apjl}{Astrophysical Journal Letters }
\newcommand{\apjs}{Astrophysical Journal Supplement }

\newcommand{\aap}{Astronomy and Astrophysics }

\newcommand{\mnras}{Mon. Not. R. Astron. Soc. }
\newcommand{\nar}{New Astronomy Reviews }
\newcommand{\nat}{Nature }

\def\keyFont{\fontsize{8}{11}\helveticabold }
\def\firstAuthorLast{Bon, N. {et~al.}} 
\def\Authors{Nata\v sa Bon\,$^{1,*}$, Edi Bon, \,$^{1}$ and Paola Marziani \,$^{2}$}
\def\Address{$^{1}$Astronomical Observatory, Belgrade, Serbia \\
$^{2}$ Osservatorio Astronomico di Padova (INAF) Padua, Italy }
\def\corrAuthor{Nata\v sa Bon}

\def\corrEmail{nbon@aob.bg.ac.rs}

\begin{document}
\onecolumn

\title[ AGN Broad Line Region variability in the context of Eigenvector 1: case of NGC 5548]{AGN Broad Line Region variability in the context of Eigenvector 1: case of NGC 5548} 

\author[\firstAuthorLast ]{\Authors} 
\address{} 
\correspondance{} 

\extraAuth{}

\maketitle

\begin{abstract}

Many  active  galactic  nuclei  (AGN)  show  strong  variability of the optical continuum.   Since  the  line flux, profile shapes and intensity ratios are changing, we  analyze  the  variability  patterns  and  possible  periodicity of Type 1 AGN NGC 5548, using the Eigenvector 1 (EV1) diagram in different variability states, taking advantage of very long term monitoring campaign  data. The preliminary results suggest that NGC 5548 - a highly variable object that over several decades has shown large amplitude continuum fluctuations and flaring behavior - remains Pop B. This means that the range in Eddington ratio, even when the source is in a bright state, remains consistent with the value of the low accreting Pop B.  We inspected EV 1 parameters of a single object though long term monitoring, assuming an inclination and black hole mass to be constant during the observational time. Our results imply that the main driver for the variations along the EV 1 diagram could be dimensionless accretion rate. If so, then it appears that the source never crossed the boundary for structural changes, indicatively placed at $L/L_{Edd} \sim$ 0.2.

\section{}

\tiny
 \keyFont{ \section{Keywords:} galaxies:active-galaxies, quasar:supermassive black holes, quasar:emission lines, line:profiles, quasar: individual (NGC 5548)} 
\end{abstract}

\section{Introduction}

\indent Differences between Type 1 and Type 2 spectra of AGNs, mainly described by the different viewing angle at the nuclear region of the galaxy, are already well known. On the other hand, there is a vast number of spectral characteristics, such as shift, width of the line, line ratios, Fe II blends, and many others that create diversity between different Type 1 spectra. One could expect that diversity also depends to some extend on the viewing angle. 

There were many efforts to systematize Type 1 spectral diversity in a parameter space called the Eigenvector 1 (EV1), that represents the linear combination of several parameters, in order to introduce some order in spectral properties. The EV1 could be seen as an equivalent to the well-known Hertzsprung-Russell diagram for stars, and therefore capable to organize Type 1 AGN into a "main sequence" of quasars. This kind of systematization allows to set observational constrains on dynamics and physical properties of broad line region. The principal component analysis of \citet{BG92} showed that there is a hidden single parameter responsible for the vast majority of spectral differences - $R_{Fe}$ - the ratio of the optical Fe II pseudo continuum to the $H\beta$ flux. This idea was later on developed by \cite{Sulentic00}, and \cite{SH2014}, among others. \\
\indent The  relation between EV1 and some theoretically motivated parameters, such as Eddington ratio, black hole mass, chemical composition, black hole spin, orientation etc., is still not clear. Most favored parameter that drives EV1 is Eddington ratio \citep{BG92,Sulentic00, 2001ApJ...558..553M, SH2014}.  \cite{2017arXiv170103694S} proposed that besides Eddington ratio the EV 1 is driven by the position of the maximum of the quasar spectral energy distribution, that is connected with the maximum of the accretion disk temperature, for the case of a \cite{SS73} accretion disk model. \cite{SH2014} argued that the viewing angle in Type 1 sources represents just a dispersion to the quasar "main sequence".\\
\indent In addition to the measurements of \cite{BG92}, \cite{Sulentic00} measured also the soft X-ray photon index and a measure of CIV $\lambda$ 1549 broad line profile velocity displacement at half maximum, in order to analyze 4D Eigenvector 1 parameter space. They showed that the "main sequence" of quasars follows some physical trends involving dimensionless accretion rate, as well as electron density which are increasing down the sequence toward strong FeII emitter, while ionization parameter is decreasing  \citep[][]{2001ApJ...558..553M}. Besides, \cite{Sulentic00} proposed a quasars dichotomy onto Pop A and Pop B according to their spectral properties. Pop B corresponds to more massive quasars \citep{Zamfir2010} and is characterized by FWHM of H$\beta$ higher than 4000 km/s and higher red asymmetry. \\
\indent With very long term monitoring campaign data, in this work we try to analyze the variability patterns on a Type 1 AGN using EV 1 diagrams in 
different variability states. We focus our analysis on the nearby and frequently observed galaxy NGC 5548, for which data from   extensive monitoring campaigns are available.

\section{Eigenvector 1 diagram for a single object multi-epoch observations }

Variability of spectra both in the continuum and in emission lines is one of the main characteristic of an active galaxy. During this time,  AGN spectra  changed slope and shape of the continuum, as well as emission line profiles (their widths and shifts) and strength  and relative intensity ratios. In a case study of a archetype of active galaxy - NGC 5548 - that has been monitored through several decades, it is possible to follow these changes, and search for a connection between spectral properties. The EV1 parameter space represents a suitable tool to analyze the AGN spectral properties though time. 

\subsection{Variability of AGN emission lines - 43 years of monitoring campaigns of NGC 5548}

Recently, \cite{Bon2016} presented the \textit{uniform} analysis of  NGC 5548 Seyfert 1 type spectra compiled from several monitoring campaigns obtained on different telescopes spanning over 43 years. Since different telescopes provide spectra with different resolution and calibration, as well as inhomogeneous aperture geometries used in different observation sets, a uniform analysis of all spectra was required. \cite{Bon2016} used ULySS - full spectrum fitting technique \citep{Koleva09,Bon2014} to calibrate the flux from all spectra in the same manner and to analyze  \textit{simultaneously} all components that contribute to the spectrum, in order to minimize dependencies between parameters of the model.  Long-term spectral variations of the continuum at 5100 $\AA$ and of the H$\beta$ line were investigated in that work. It was found that the light and radial velocity curves show periodic variation with the periodicity of nearly 16 years. Also, NGC 5548 was noticed before to be a changing look AGN \cite[see for e.g.][]{2007ApJ...668..708S}, with the clear appearance and disappearance of broad H$\beta$ component through time: the spectral type changed from Seyfert 1 to Seyfert 1.8. \\
\indent Using the same technique as presented in \cite{Bon2016}, here we measured 
EV1 diagram properties of NGC5548 spectra - FWHM(H$\beta$) versus R$_{Fe}$, 
in order to investigate the problem of the physical properties of AGN variability along the QSO "main sequence". Modeling of emission lines, and the contribution of the continua of an AGN and host galaxy was obtained as described in \cite{Bon2016}, while Fe II pseudo continuum was modeled with the template described in \cite{2009A&A...495...83M}.\\

\begin{figure}[h!]
\begin{center}
\includegraphics[width=15cm]{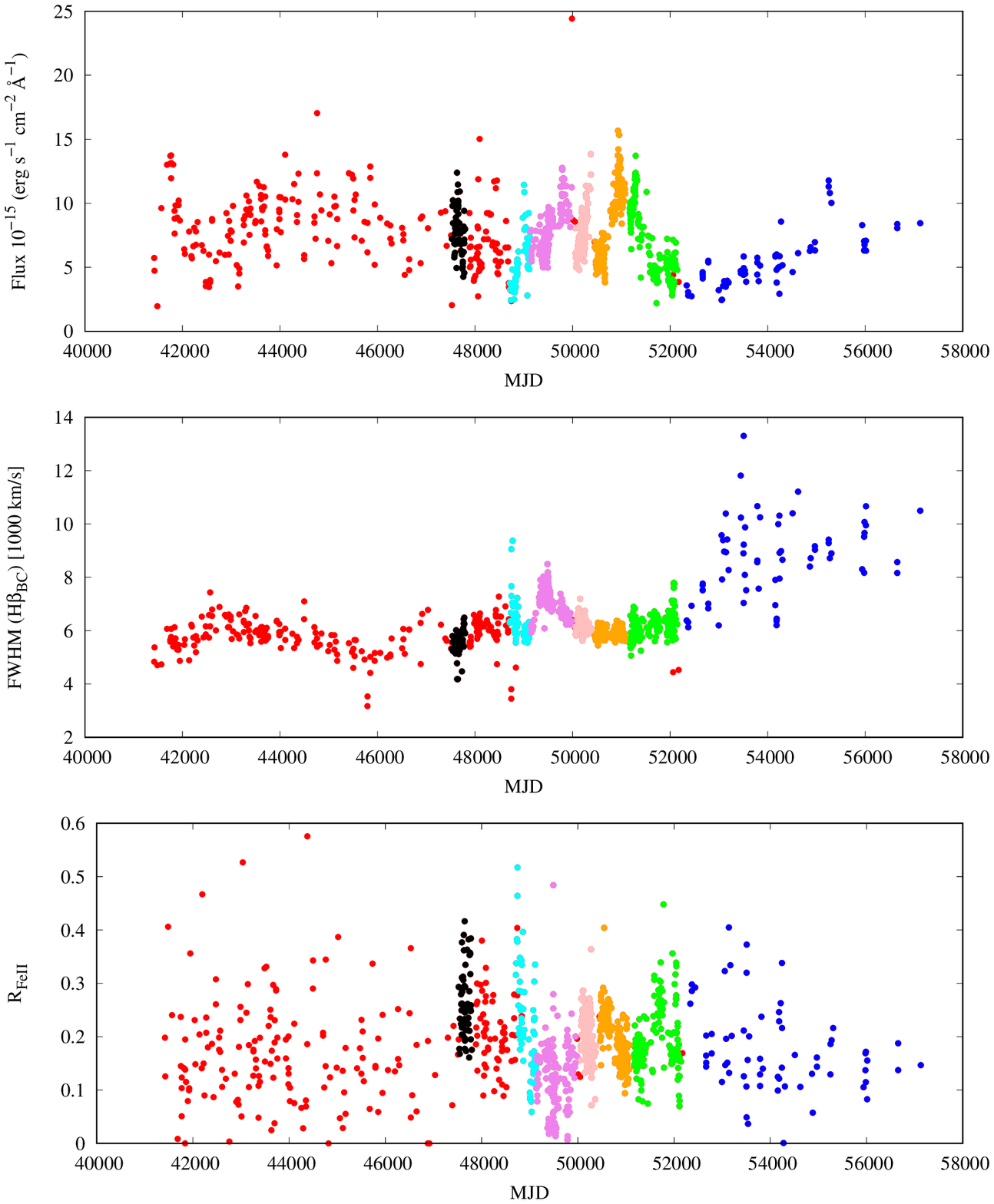}
\end{center}
\caption{Variability in the NGC 5548 spectra: (top) The light curve of the continuum measured at \\
	5100 $\AA$; (middle) FWHM H$\beta$ variations; (bottom) $R_{FeII}$ change with the time. Colors correspond to different time intervals.}\label{fig:Light_curve}
\end{figure}

\begin{figure}[h!]
	\begin{center}
		\includegraphics[width=15cm]{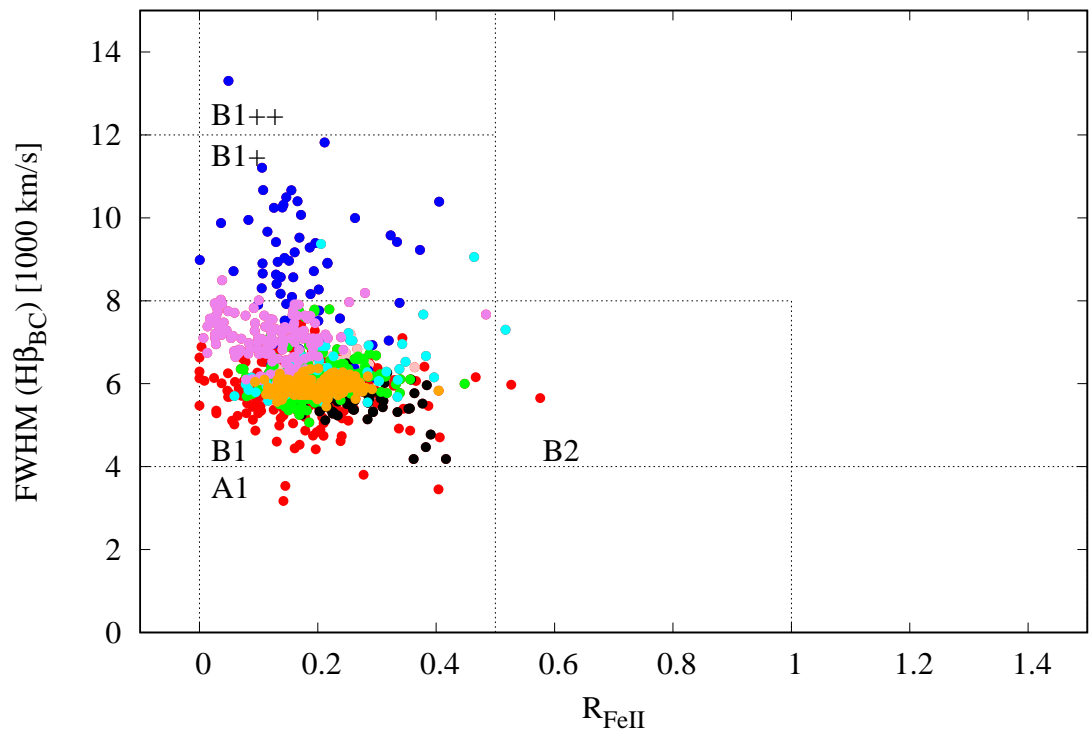}
	\end{center}
	\caption{The EV1 diagram of NGC 5548 spectral properties during 43 years of monitoring campaigns. Colors correspond to those in the Fig. \ref{fig:Light_curve}.}\label{fig:EV1}
\end{figure}

\begin{figure}
	\begin{center}
		\includegraphics[width=0.4\linewidth]{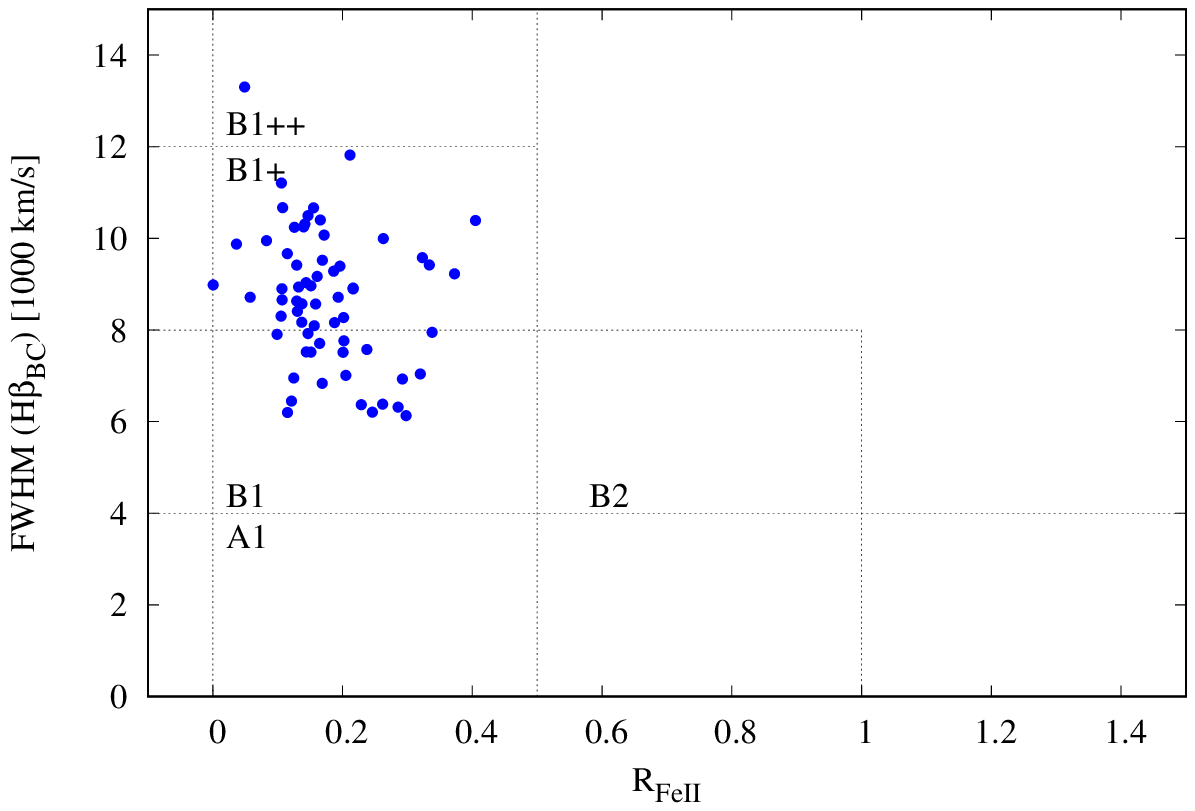}
		\includegraphics[width=0.4\linewidth]{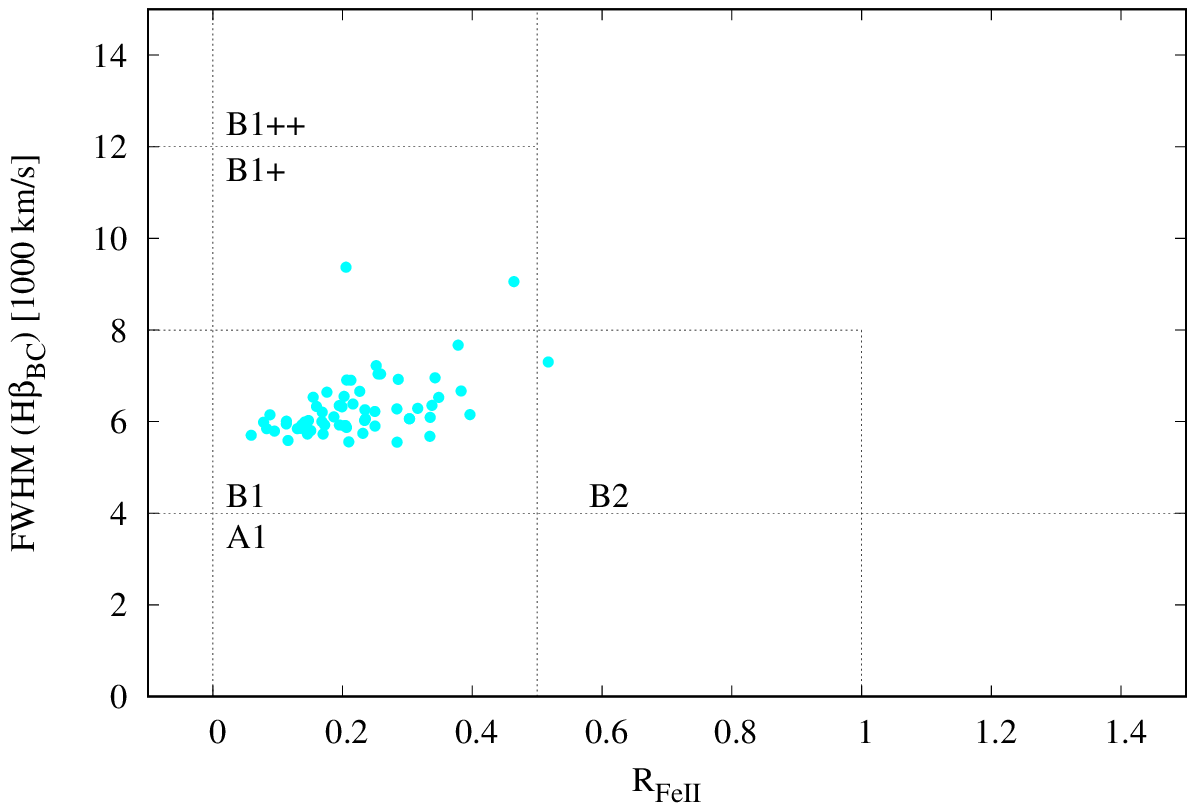}
		\includegraphics[width=0.4\linewidth]{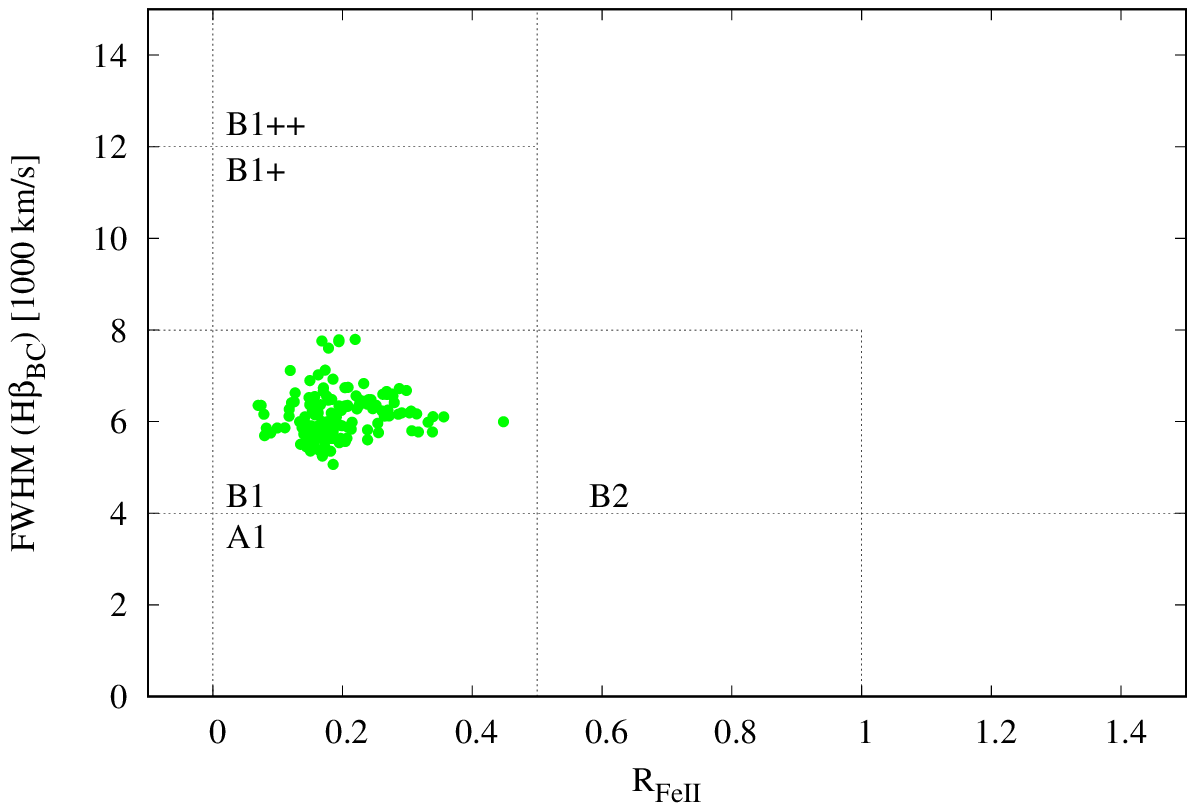}
		\includegraphics[width=0.4\linewidth]{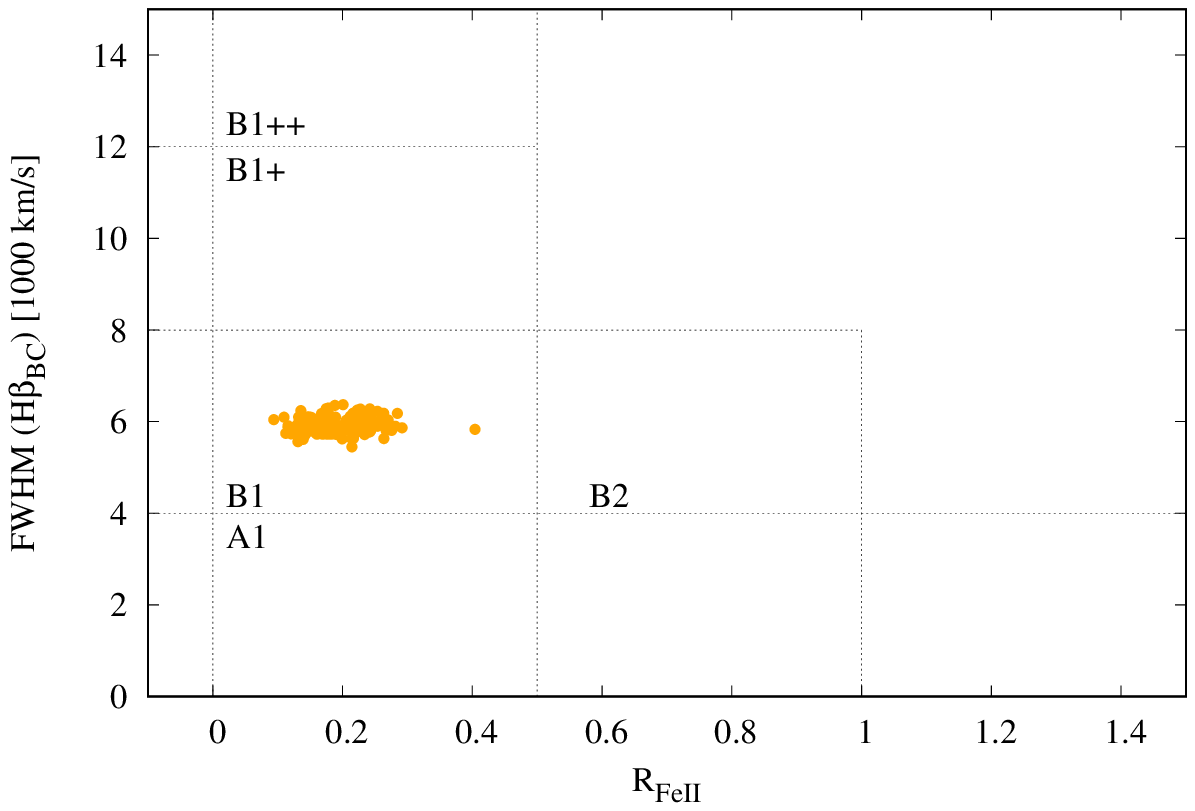}
		\includegraphics[width=0.4\linewidth]{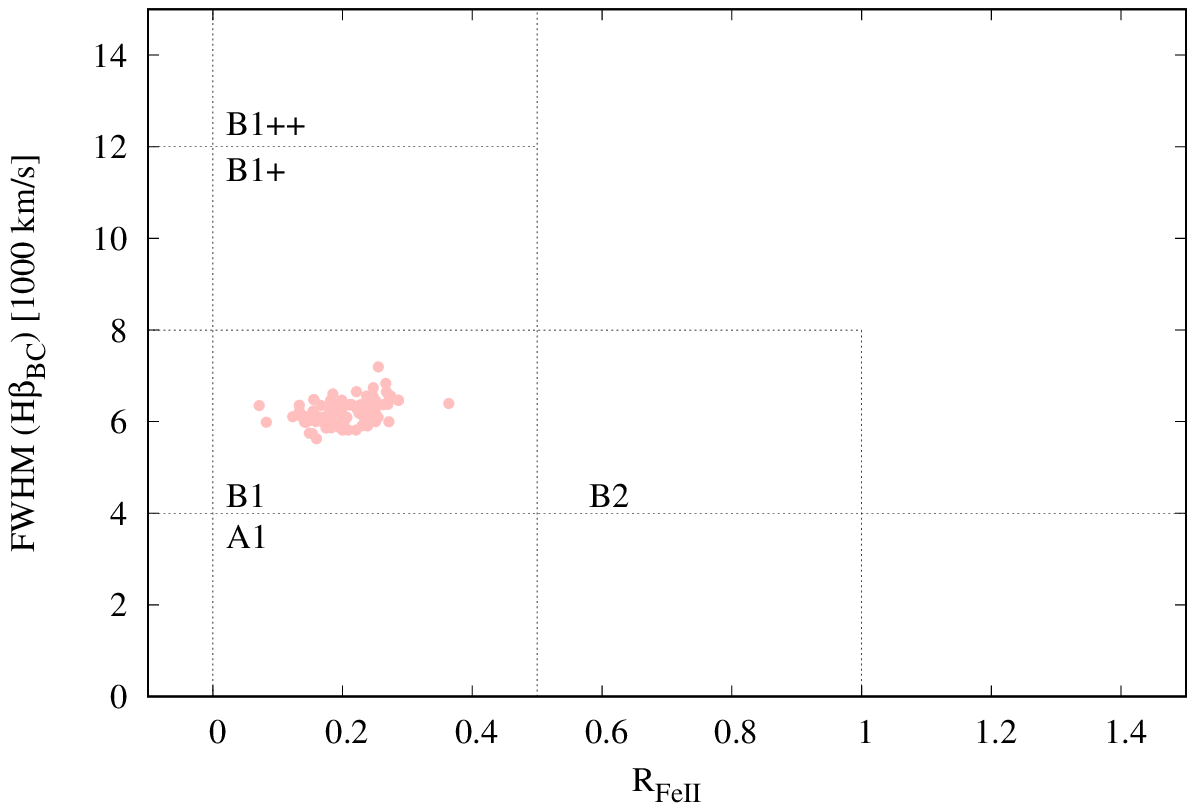}
		\includegraphics[width=0.4\linewidth]{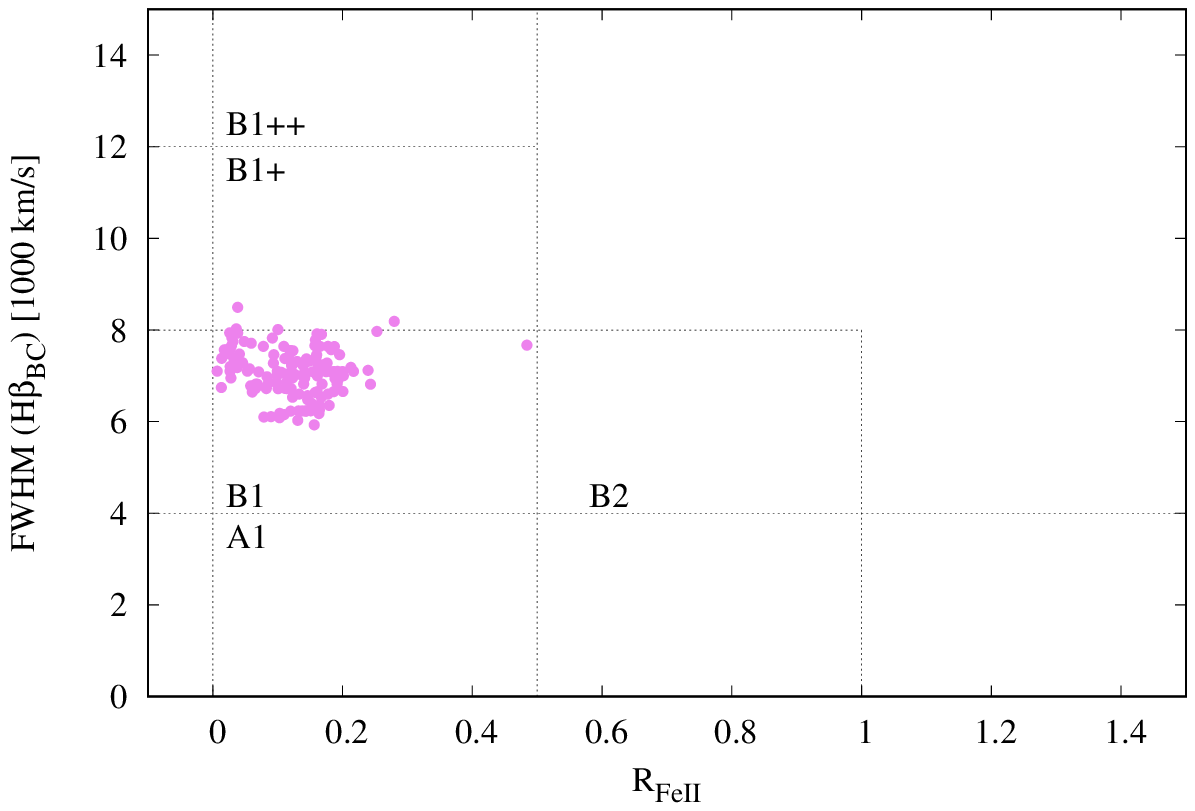}
		\includegraphics[width=0.4\linewidth]{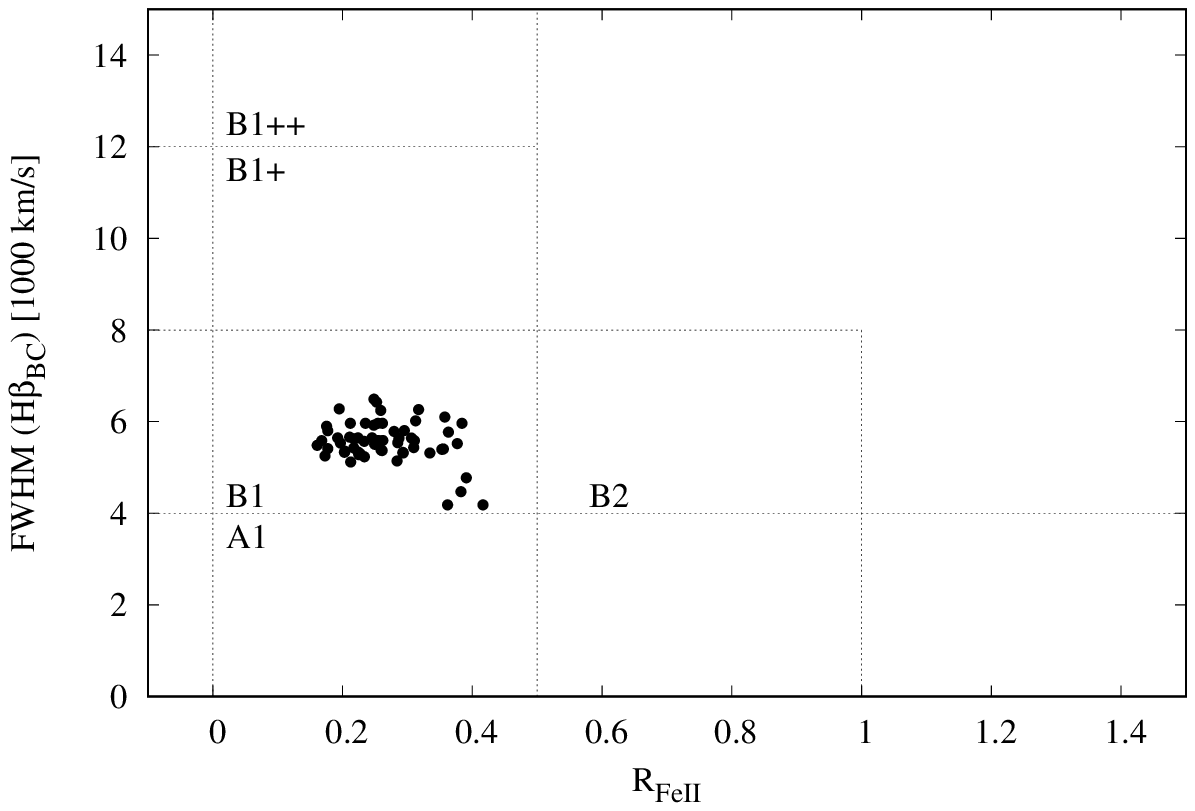}
	\end{center}
	\caption{The EV1 diagram of NGC 5548 spectral properties for different time intervals. Colors correspond to those in the Fig. \ref{fig:Light_curve}.}\label{fig:EV1_colors}
\end{figure}

\indent Figure \ref{fig:Light_curve} shows (a) the light curve in the continuum measured at 5100 $\AA$, (b) the FWHM  of H$\beta$ variations during the time, as well as (3) the variability of the $R_{FeII}$ with the time. We also analyzed some fast changing flux variations with flare-like behavior. We selected different time intervals (shorter and longer, in high state and low state of activity, as well) in order to analyze their EV1 properties. We presented these intervals with different types of variations in different colors, where each color in all plots correspond to the same time interval, in order to present measured parameters on the EV1 diagram in the Fig. \ref{fig:EV1} in colors that correspond to those in the Fig. \ref{fig:Light_curve}. One can notice that NGC 5548 spectra in an extreme low state of activity (transition from Type 1 to Type 1.8), changes from Pop B1 to Pop B1+ (and in same cases even to Pop B1++), while in a high state object changes towards Pop B1 (and in few cases even to Pop A1). It means that NGC 5548 changes, but mainly stays Pop B. The data of the paper are meant to cover a time lapse that is several time the dynamical time scale.  Fig. \ref{fig:EV1} shows the variability  of a single AGN through a large time interval which fills the area of a whole population of AGN with similar observational characteristics (in this case Pop B).  At the same time, we are not expecting to find the same relation that are found in reverberation mapping campaign (RM \citep{2002ApJ...581..197P}, and references therein)  on shorter time scales and with frequent sampling. In the Fig. \ref{fig:EV1_colors} we analysed short and long term variations separately on EV 1 diagrams. As one can see the behavior for each segment on time line
is very different. The biggest structural change we noticed in the time interval colored in blue and cyan.  \\

\begin{figure}
	\begin{center}
		\includegraphics[width=15cm]{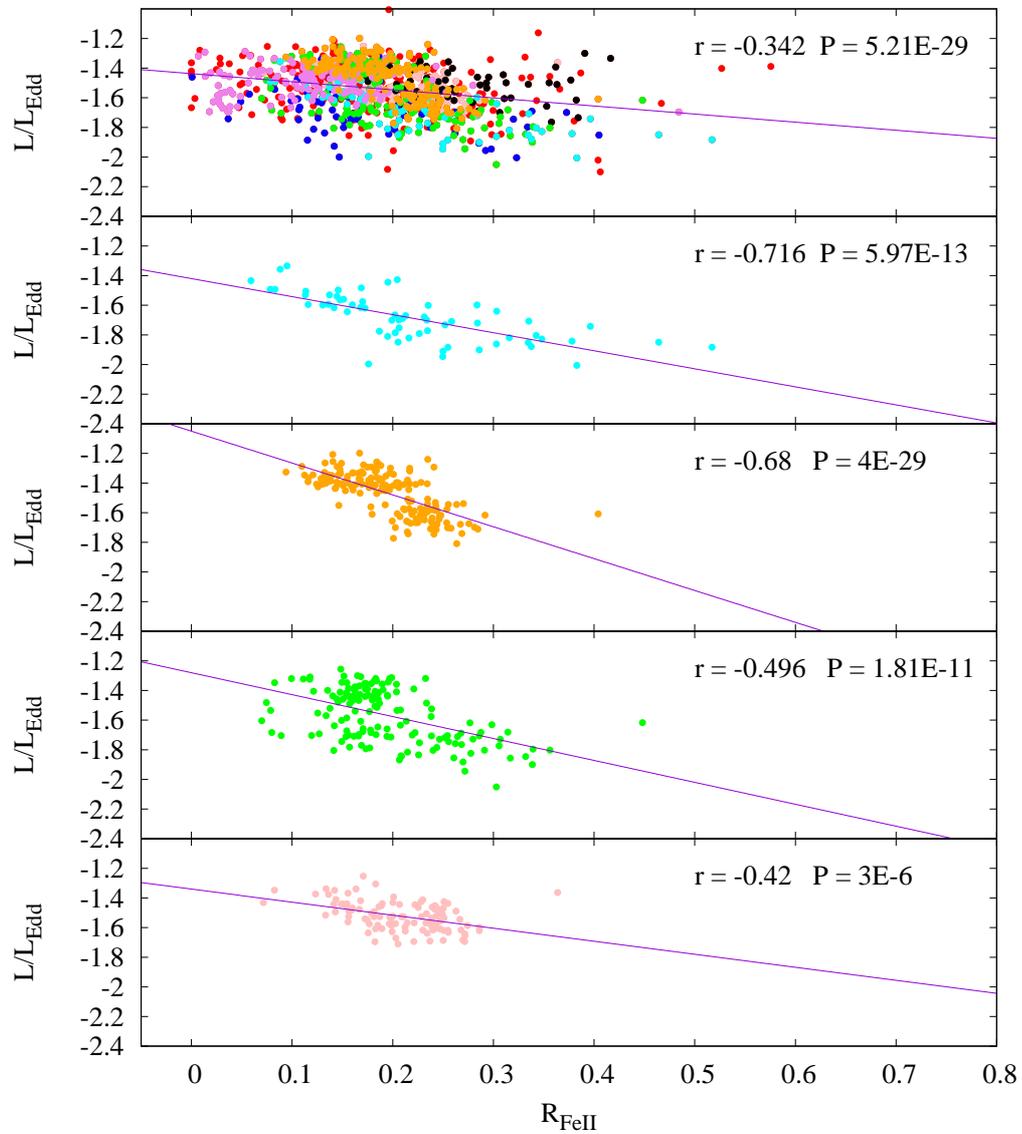}\\
	\end{center}
	\caption{The negative trend: $L/L_{Edd}$ decrease with the increase of $R_{FeII}$ with the time. The strongest anti-correlation was found for well defined time intervals presented on the middle and bottom panel. Colors correspond to those in Fig. \ref{fig:Light_curve} and Fig. \ref{fig:EV1}.}\label{fig:AC_multiplot}
\end{figure}

\begin{figure}
	\begin{center}
		\includegraphics[width=15cm]{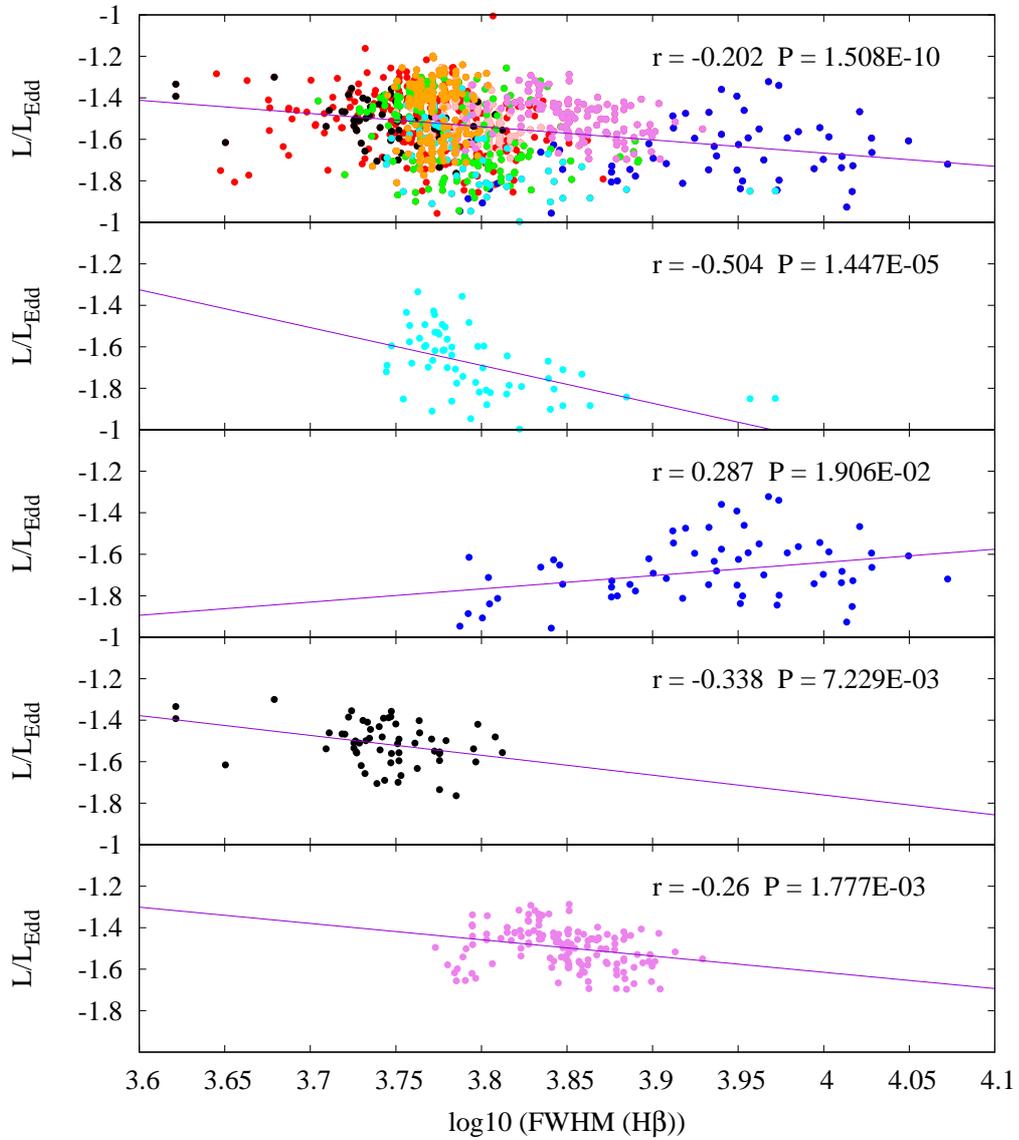}\\
	\end{center}
	\caption{ $L/L_{Edd}$ vs $log_{10}(FWHM)$ trends for long term (top panel) and short term variations (mid and bottom panels). Correlation coefficients are marked on each plot. Colors correspond to those presented in Fig. \ref{fig:Light_curve} and Fig. \ref{fig:EV1}.}\label{fig:AC_multiplotFWHM}
\end{figure}

\indent It is expected that both inclination and the black hole mass play a significant role in the physical characterization of the AGN main sequence diagram, but in the case of the inspection of EV1 parameters of a single object through time, we expect inclination and black hole mass to be  constant during the whole observational time. Therefore, the main driver of the variability along the EV1 diagram is expected to be associated with variations in accretion rate (since the black hole mass is fixed, accretion rate is proportional to both luminosity [for fixed radiative efficiency] and Eddington ratio). To analyze that, we show  variations of $L/L_{Edd}$ against $R_{FeII}$ in Fig. \ref{fig:AC_multiplot}. We notice that decrease of $L/L_{Edd}$ is followed by an increasing $R_{FeII}$, mainly in intervals when the flux shows large changes in very short time intervals (represented with high Pearson's correlation coefficient), both in the continuum, as well as in the H$\beta$ line, indicating that changes along the EV1 diagram of the single object could be due sudden and fast changes in accretion rate. We find also that accretion rate and FWHM  show modest  correlation coefficient in complete monitoring interval, while in short time variations it may be quite high (see Fig. \ref{fig:AC_multiplotFWHM}). The $L/L_{Edd}$ is obtained from the luminosity and the black hole mass, but since the mass is assumed constant, then the FWHM should not affect the $L/L_{Edd}$. We calculated $L/L_{Edd}$ using continuum flux measurements at 5100 $\rm \AA$ 
assuming the bolometric correction factor 10 to the specific luminosity measured at 5100 $\rm \AA$ \citep[see,][]{Sulentic2006}. The mass of the black hole is assumed to be $5.7 \times 10^{7} \  M\odot$, as in the paper \citep{Bon2016}. Therefore, the study of a single object has the advantage that FWHM  and  $L/L_{Edd}$ measurements are independent. On the converse, in the study of quasar  samples, the FWHM is used to compute $M_{BH}$, and hence $L/L_{Edd}$ and FWHM are not independent parameter. \\


\section{Discussion and conclusion}

\indent Since NGC 5548 has not undergone major structural change over $\sim$ 40 years of observations, if the physical parameter that drive EV 1 is Eddington ratio, then it appears that the source never crossed the boundary for structural changes, indicatively placed at $L/L_{Edd}$ $\sim$ 0.2.

\begin{figure}
	\begin{center}
		\includegraphics[width=15cm]{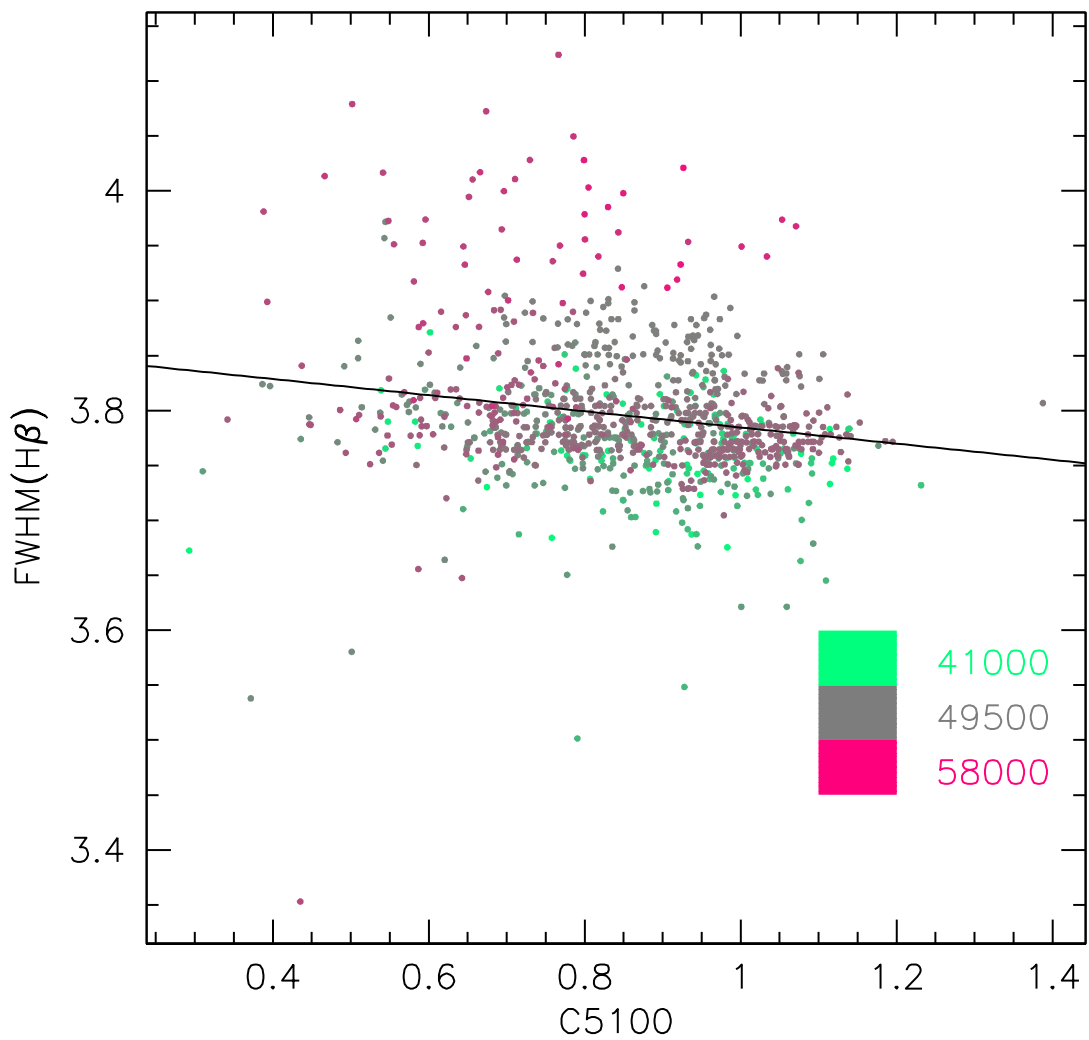}\\
	\end{center}
	\caption{Anti-correlation between FWHM of H$\beta$ and continuum specific flux at 5100 \AA. The line shows an unweighted least squares fit. Colors of data points are coded from green to magenta according to Julian date, from  2441000 to 2458000. 
		}\label{fig:Paola}
\end{figure}

\begin{figure}
	\begin{center}
		\includegraphics[width=15cm]{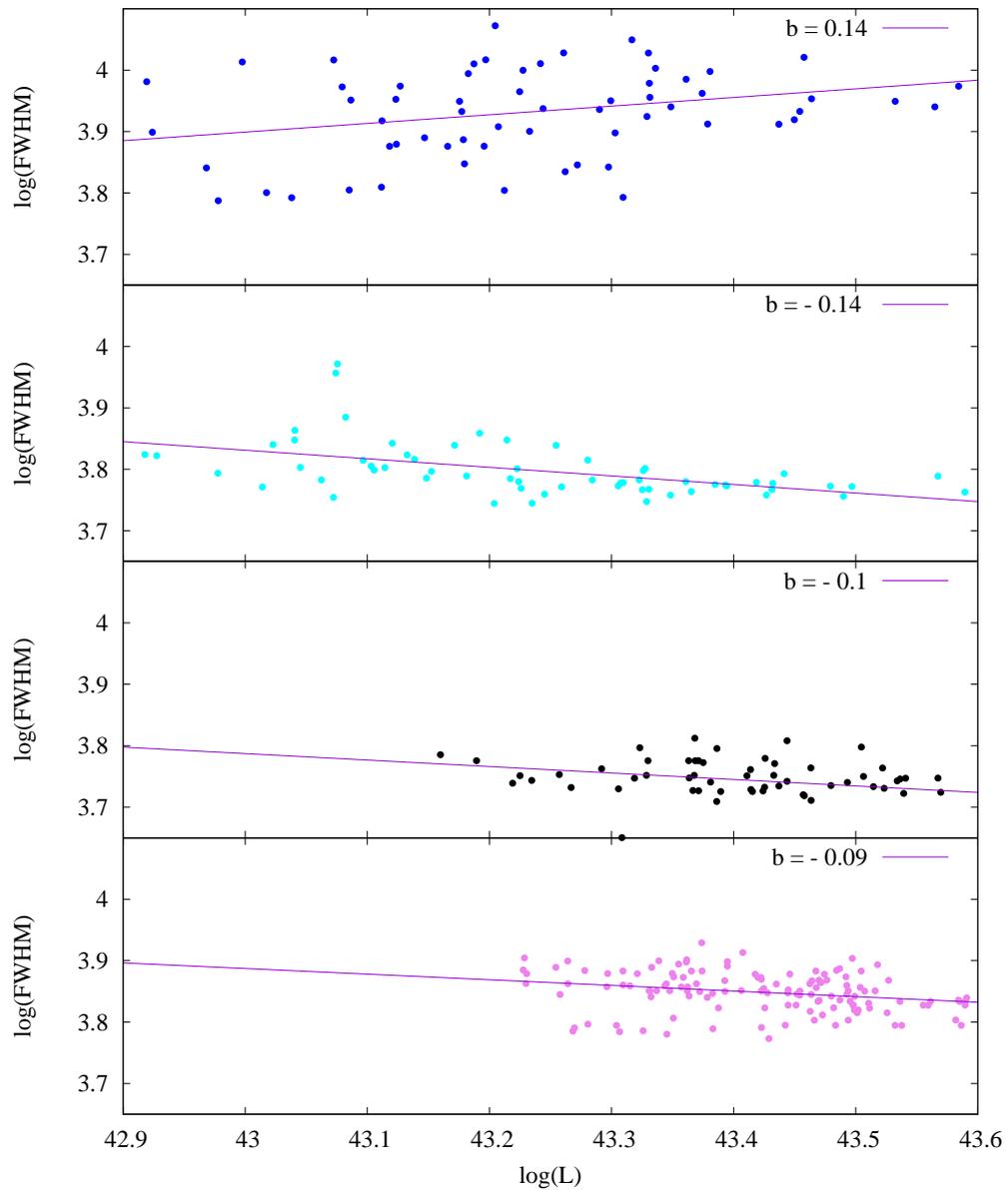}\\
	\end{center}
	\caption{\rm Trends of luminosity against FWHM of H$\beta$ for long term (blue represents last 12 years of observations) and short term variations (cyan, black and pink are order of months).
		Lines shows an unweighted least squares fit. Colors of data points correspond to those presented in the Fig. \ref{fig:Light_curve}.
		Values of slope trends are marked on each plot.
	}\label{fig:LvsFW}
\end{figure}

A decrease in FWHM with increasing $L/L_{Edd}$\ would be consistent with the expectations of an increasing effect of radiation forces with increasing luminosity. The absence of a strong relation is only in apparent contradiction with the expectation of the model by \citet{2010ApJ...724..318N}. We found a very weak anti-correlation between FWHM and $L/L_{Edd}$, with a slope of $\approx -0.07$ \footnote{$log FWHM \sim (-0.073 \pm 0.0125) (log L) + ( 3.858 \pm 0.0108 )$} (with Pearson correlation coefficient of -0.18, and a significance of $7 \times 10^{-9}$). We also found a significant but weak anticorrelation between FWHM of H$\beta$ line and continuum, with a modest slope (see Fig. \ref{fig:Paola}). For the typical $L/L_{Edd}$ of NGC 5548, radiation forces have a relatively little effect on the dynamics of the line emitting clouds. A FWHM change should be limited to $\lesssim 10$\% (Table 1 of \citealt{2010ApJ...724..318N}), as indeed suggested by the trend found from  the actual data (with a slope $-0.07$, a three-fold  increase in continuum implies a narrowing by $\approx 8$\%). There is therefore no contradiction between a weak and shallow correlation   and a minor role of radiation forces. \\
\indent In the studies of quasar samples it is shown that FWHM of H$\beta$ increases, because of the increase of the black hole mass 
For an individual object the story is different. The most reasonable assumption is that the virial product is constant $r  \times FWHM^{2} =$ constant\footnote{Also f, the form factor should not be constant rigorously speaking, but let us assume at the moment it is}. If r scales with luminosity as $r \propto L^a$, then $FWHM \propto L^{-a/2}$,
which is not far from the trend we detect for the complete monitoring interval (see Fig. \ref{fig:Paola}). But for short time variability epochs trends are different for each segment that we defined (see Fig. \ref{fig:LvsFW} and marked trends on each panel). It is interesting that in the time interval marked with blue, we see the positive trend of luminosity against FWHM of H$\beta$ broad emission line  (see Fig. \ref{fig:LvsFW}). This time interval corresponds to long term variation of flux slowly increasing for 12 year, starting from deep minimum in the low state and ending up towards the high state. In contrary to it, short term variations (for e. g., cyan and pink) shows negative trends (see panels in Fig. \ref{fig:LvsFW}).

In addition, right the recent work of 
\citet{,Pei2017ApJ} finds, for two different time lapses T1 and T2, mean continuum fluxes at 5100 \AA\ and line H$\beta$ line widths: T1) mean flux = 11.31 $\pm$ 0.08, FWHM (km $s^{-1}$) = 9612 $\pm$ 427 and
T2)  mean flux = 12.51 $\pm$ 0.04,  FWHM (km s$^{-1}$) = 9380 $\pm$ 158.
The trend between the two time ranges, implies a decrease in the FWHM when the source is brighter. Without considering the measurement dispersion in FWHM the slope is right -0.25, as predicted by the elementary consideration above in case of $r \propto L^{1/2}$, and therefore $FWHM \propto L^{-1/4}$, since  $r  \propto FWHM^{-2}$.

The resulting slope for the full sample (-0.07) is the result of mixing together different epochs in which the response of the BLR is different as per the short-term “breathing effect” first described by 
\cite{NezerMaoz1990ApJ}
 and discussed in full by \cite{Korista2004ApJ}.


We believe that the weak anticorrelation we found deserves further analysis  separately considering different states of continuum level/behavior, as well as changes in the structure factor f$_S$ that, evidently, cannot be assumed as a constant, but we do not think that the shallow trend is inconsistent with existing data of reverberation mapping campaigns.


 Flattened, disk like, structure combined with surrounding isotropic region is usually good approximation of the BLR structure, so the line profile of population B sources can be approximated as the sum of a accretion disk profile component 
 describing wings of the broad emission line, + an isotropic component 
 describing the intermediate broad core of the broad line which is blending the double peaked shape of an accretion disk profile, producing the single peaked line profile with very broadened wings, typically seen in the AGN broad emission line profiles 
 \cite[see, e. g.][]{Pop2004,Bon2008SerAJ,Bon2009NewAR,bonetal09}. The
optically thick core responding more strongly than continuum changes that the accretion disk component could also produce an anti-correlation between FWHM and continuum intensity. We plan to test this possibility in a forthcoming work.

The trend  $L/L_{Edd}$ against $R_{FeII}$ is also difficult to explain. In principle, both \citet{2001ApJ...558..553M} and \citet{SH2014} agree that an increase in $L/L_{Edd}$ should lead to higher $R_{FeII}$. This deduction was however reached from the analysis of a large sample of quasars, and not from the behavior of an individual source. Since  $R_{FeII}$  is the ratio of two quantities that  both vary, it is important to know how the intensity and the EW of H$\beta$\ and FeII vary separately.  H$\beta$\  
shows a highly significant anti-correlation between its equivalent width and the continuum, with a lsq best-fit slope of $\approx -0.26$ which is highly significant for 980 data points. This anti-correlation (the ``Baldwin effect'') is significantly steeper for FeII: the slope is $\approx -0.60$.  These dependencies reflect a different response to continuum changes for H$\beta$ and FeII: while the flux of  H$\beta$\ is correlated with continuum flux with a slope $\approx$ 0.80, implying a strong response, the response of FeII is much weaker, with I(FeII) $\propto$ 0.40 I(cont). Therefore, the EW of the FeII increases more than the EW of H$\beta$\ when continuum is low, in turn increasing  $R_{FeII}$. These considerations suggest that a parallel between the behavior of large samples and the one associated with the variability of an individual object cannot be drawn, and that NGC 5548 does not challenge the notion of a positive relation between 
$R_{FeII}$ and $L/L_{Edd}$ found in large quasar sample.

The preliminary results summarized in this paper suggest that NGC 5548 -- a highly variable object that over several decades has seen large-amplitude continuum fluctuations and flaring behavior  -- remains a source of Pop. B,  its variability notwithstanding. The location  of NGC 5548 in the optical plane of E1 remains constrained within spectral type B1.  The range in Eddington ratio, even when the source is in a bright state, remains consistent with the value of the lowly - accreting Pop. B, implying that the source has not undergone major structural changes  during the 40+ years it was monitored.      

\section*{Conflict of Interest Statement}
The authors declare that the research was conducted in the absence of any commercial or financial relationships that could be construed as a potential conflict of interest.
 
\section*{Author Contributions}
EB is responsible for developing the idea, NB for doing the fitting, NB, EB and PM for analysing results, writing the text of the manuscript, and for discussing results.

\section*{Funding}
This research is part of projects 
176003 “Gravitation and the large scale structure of the
Universe” and 176001 “Astrophysical spectroscopy of extragalactic
objects” supported by the Ministry of Education and
Science of the Republic of Serbia. 

\section*{Acknowledgments}
We would like to thank to Jack Sulentic and Martin Gaskell for helpful comments.
This research is part of projects 
176003 “Gravitation and the large scale structure of the
Universe” and 176001 “Astrophysical spectroscopy of extragalactic
objects” supported by the Ministry of Education and Science of the Republic of Serbia. 





\section*{Figure captions}


\end{document}